\documentclass[]{aastex701}

\usepackage{amsmath}
\usepackage{bm}

\begin{document}

\title{Formation of a Magnetic Flux Rope Prior to the Eruption: Insight from a Radiative MHD Simulation of Active Region Emergence}

\author[orcid=0000-0002-6799-4340,gname=Can, sname=Wang]{Can Wang}
\affiliation{School of Astronomy and Space Science, Nanjing University, Nanjing 210023, People's Republic of China}
\affiliation{Key Laboratory of Modern Astronomy and Astrophysics (Nanjing University), Ministry of Education, Nanjing 210023, People's Republic of China}
\affiliation{Astronomical Observatory, Kyoto University, Sakyo-ku, Kyoto 606-8502, Japan}
\email{canw@smail.nju.edu.cn}  

\author[orcid=0000-0001-5457-4999,gname=Takaaki, sname=Yokoyama]{Takaaki Yokoyama} 
\affiliation{Astronomical Observatory, Kyoto University, Sakyo-ku, Kyoto 606-8502, Japan}
\email{yokoyama.takaaki.2a@kyoto-u.ac.jp}

\correspondingauthor{Feng Chen}
\email{chenfeng@nju.edu.cn}
\author[orcid=0000-0002-1963-5319,gname=Feng,sname=Chen]{Feng Chen}
\affiliation{School of Astronomy and Space Science, Nanjing University, Nanjing 210023, People's Republic of China}
\affiliation{Key Laboratory of Modern Astronomy and Astrophysics (Nanjing University), Ministry of Education, Nanjing 210023, People's Republic of China}
\email{chenfeng@nju.edu.cn}

\author[0000-0002-6550-1522,gname=Chen,sname=Xing]{Chen Xing}
\affiliation{School of Astronomy and Space Science, Nanjing University, Nanjing 210023, People's Republic of China}
\affiliation{Key Laboratory of Modern Astronomy and Astrophysics (Nanjing University), Ministry of Education, Nanjing 210023, People's Republic of China}
\email{chenxing@nju.edu.cn}

\author[0000-0002-4978-4972,gname=Mingde,sname=Ding]{Mingde Ding}
\affiliation{School of Astronomy and Space Science, Nanjing University, Nanjing 210023, People's Republic of China}
\affiliation{Key Laboratory of Modern Astronomy and Astrophysics (Nanjing University), Ministry of Education, Nanjing 210023, People's Republic of China}
\email{dmd@nju.edu.cn}

\author[0000-0001-7961-7617,gname=Zekun,sname=Lu]{Zekun Lu}
\affiliation{School of Astronomy and Space Science, Nanjing University, Nanjing 210023, People's Republic of China}
\affiliation{Key Laboratory of Modern Astronomy and Astrophysics (Nanjing University), Ministry of Education, Nanjing 210023, People's Republic of China}
\affiliation{Institute for Space-Earth Environmental Research, Nagoya University, Chikusa-ku, Nagoya 464-8601, Japan}
\email{zekunlu@smail.nju.edu.cn}

\begin{abstract}
Magnetic flux ropes (MFRs) are fundamental magnetic structures in solar eruptions, whose formation is generally attributed to (1) the emergence of subsurface flux tubes or (2) flux cancellation driven by photospheric horizontal flows and magnetic reconnection. Both mechanisms can operate simultaneously during active region evolution, making their relative contributions challenging to quantify. Here, we analyze the formation of a flux rope in a MURaM radiative magnetohydrodynamic (RMHD) simulation, which formed and evolved for approximately three hours before an M-class flare. The formation process is quantified by magnetic helicity flux, which drives the non-potential evolution of magnetic field, with its advection and shear terms on the photosphere corresponding to the emergence and photospheric horizontal flows, respectively. Examining the helicity injected into the flux rope through the photosphere, we find both terms increase significantly as the eruption approaches, with the shear term prevailing overall. Height-dependent analysis of helicity flux, together with magnetic field and velocity distributions, further reveals a gradual transition from the shear to the advection term with an increasing altitude, which is driven by magnetic reconnection above the photosphere. Our results provide quantitative evidence that flux cancellation governs flux rope formation, arising naturally from magnetic field reorganization during active region evolution: as flux emergence transports magnetic flux upward, photospheric shearing motions adjust magnetic field and inject helicity into solar atmosphere, and magnetic reconnection ultimately assembles the main body of flux ropes.
\end{abstract}


\section{Introduction} \label{sec:intro}
Magnetic flux ropes (MFRs) are the key magnetic structures involved in solar eruptions. According to the standard flare model, as a flux rope rises into the higher corona, magnetic reconnection occurs within the current sheet beneath it. The magnetic energy released during reconnection contributes to local plasma heating and particle acceleration, subsequently heating the lower atmosphere via thermal conduction and energetic electron beams, ultimately producing flares. Meanwhile, part of the magnetic energy is converted into the kinetic energy of the rising flux rope, establishing positive feedback with its escape and contributing to its development into a coronal mass ejection (CME) \citep{Lin2000,Shibata2011,Aulanier2012,Aulanier2013,Janvier2013,Janvier2014}. 

Qualitatively, flux ropes can be identified by two key characteristics. First, magnetic field lines within a flux rope exhibit a coherent helical structure, winding around a common axis with consistent morphological features. Second, they show distinct magnetic connectivity compared with surrounding magnetic structure. In situ measurements of interplanetary CMEs have uncovered large-scale twisted magnetic structures and their correlation with coronal flux ropes \citep{Gopalswamy2018}, providing direct evidence for the role of flux ropes in solar eruptions. While most studies have focused on flux rope dynamics during eruptions, understanding their long-term evolution prior to eruption—especially the formation process—is also crucial for understanding and predicting solar eruptions.

Proposed formation mechanisms of MFRs generally fall into two major categories. The first emphasizes the role of partly or bodily emergence of flux tubes from the solar interior. By inserting twisted flux tubes beneath the photosphere and initiating their emergence via electric field, buoyancy forces, or velocity perturbations, numerical simulations have successfully reproduced the injection of non-potential energy into solar atmosphere, as well as the subsequent formation and even eruption of MFRs \citep{Manchester2004,Fan2004,Archontis2008,Hood2009,Zhuleku2025}. Observational evidence also supports the scenario of emerging twisted flux tubes, such as the so-called `sliding doors’
effect \citep{Okamoto2008,Lites2010} and the evolution of magnetic tongues appearing as elongated polarities on the photosphere \citep{Luoni2011}. Despite ongoing debates about the existence of twisted flux ropes in the convection zone and challenges to the coherent emergence of large-scale magnetic structure, it is established that the twisted magnetic structures stored in the convection zone can be transported into solar atmosphere through flux emergence, as indicated by the advection term of magnetic helicity or winding flux \citep{MacTaggart2021}. In this paper, the term `flux emergence mechanism' is used in a broad sense, referring to the process by which twisted magnetic structures in the convection zone rise through the photosphere into the solar atmosphere, without strictly requiring a fully coherent flux tube below the surface. 

The second category focuses on the occurrence of flux cancellation. While the emergence of U-loops can also manifest as flux cancellation, it is more appropriately classified under the emergence mechanisms discussed above. In contrast, the classical flux cancellation model for flux rope formation emphasizes the role of photospheric horizontal motions, especially the shearing and the converging motions, and magnetic reconnection \citep{vanBallegooijen1989}. In observations, the simultaneous occurrence of sigmoids or filament development and the cancellation of opposite magnetic flux near the polarity inversion line (PIL) is often accompanied by shearing and/or converging motions of polarities, as well as brightenings that indicate the occurrence of reconnection \citep{Chae2001,Green2011,Yardley2016,Dai2022,Li2023}. Many numerical simulations also successfully reproduce the formation of flux ropes, under the flux cancellation framework by adopting photospheric horizontal flows and magnetic reconnection \citep{Amari2003,Mackay2006,Aulanier2010,Xia2014,Guo2024,Xing2024a,Xing2025} . 

While theoretical models are well-established, significant challenges still exist in understanding the formation process of flux ropes, especially in distinguishing the role of each mechanism in detail. On the one hand, the lack of direct observations in convection zone and the limitations of remote-sensing observations in solar atmosphere result in ambiguous detections of both flux emergence and reconnection. Consequently, most studies only qualitatively discuss the contributions of these processes without quantitative analysis \citep{Liu2019,Zheng2020}. Magnetic helicity flux on the photosphere can be decomposed into a shear term and an advection term, which quantify the contributions of horizontal motions and direct emergence, respectively. Although a few studies have investigated helicity flux and the contributions of different terms, they mainly analyzed the entire active region instead of the flux rope itself \citep{Wang2018,Sun2024}, which makes it difficult to extract information specific to the flux rope. On the other hand, the simplified treatment of energy process and flux emergence in simulations introduces biases when compared with the self-consistent formation of flux ropes during the evolution of active regions on the real Sun. Therefore, numerical simulations that incorporate a more realistic treatment of energy transport and flux emergence, such as Bifrost \citep{Gudiksen2011}, MURaM \citep{vogler2005,Rempel2017}, R2D2 \citep{Hotta2019}, and others, could provide a promising approach for a deeper understanding of flux rope formation.

In this paper, we analyze the simulation of a complex active region (AR) conducted with the MURaM code \citep{Chen2022}. Besides the adoption of radiative transfer and a tabular equation-of-state in the MURaM, which enables the reproduction of a more realistic convective environment, the simulation by \citet{Chen2022} employs the data from a dynamo simulation as input for flux emergence, generating long-lasting, complex AR and the self-consistent photospheric velocity fields coupled with AR evolution. During the evolution lasting approximately 48 solar hours, the simulated active region gives rise to more than 100 eruptive events and diverse phenomena similar to those on the real Sun, such as confined eruptions \citep{Wang2022,Wang2023}, large-scale extreme ultraviolet (EUV) waves \citep{Wang2021}, and super-hot corona loops \citep{Lu2024}. \citet{Chen2023} analyzed the most energetic event in the simulation, revealing the existence of a pre-eruptive magnetic flux rope and examining its evolution during the eruption. Here, we focus specifically on the flux rope’s formation process, tracing its development from the initial flux rope seed until eruption onset by using the same dataset of \citet{Chen2023}. By examining the temporal variation and spatial distribution of helicity flux, we conduct quantitative analysis of the roles of the flux emergence and the flux cancellation mechanisms, providing new insights into the formation of flux ropes as the active regions evolves.

The paper is organized as follows. Section \ref{Setup} describes the simulation setup and the dataset. Section \ref{sec:result} presents the main results, including the identification of flux rope and the analysis of magnetic helicity flux. Finally, we discuss and summarize this work in Section \ref{sec:discussion}.

\section{Numerical Simulation Setup and Data}\label{Setup}
The simulation analyzed in this study is performed in a Cartesian coordinate system with a domain size of $L_x \times L_y \times L_z = 196.6$ Mm $\times$ 196.6 Mm $\times$ 122.9 Mm. Along $z$-direction, the layer where horizontally averaged optical depth approximately equals unity is marked as $z = 0$ Mm and regarded as the photosphere. Thus, the simulation covers a height range from the upper convection zone at about $-$9.6 Mm to more than 100 Mm in the corona. The calculation domain is resolved by uniform grids of $N_x \times N_y \times N_z = 1024$ $\times$ 1024 $\times$ 1920, yielding a spatial resolution of $\Delta x = \Delta y = 192$ km and $\Delta z = 64$ km. Periodic boundary conditions are applied in the lateral directions for all variables. Although the periodic boundary conditions may alter the large-scale magnetic topology and the amount of energy accumulated within the simulation box, the pre-eruptive flux rope is compact and well-separated from the boundaries, and the evolution of magnetic energy is not significantly affected by the boundary setup \citep{cheung2019}. At the top boundary, the horizontal velocity and thermal variables are set to be symmetric, while the vertical velocity is strongly damped to ensure numerical stability. The magnetic field in the ghost cells is specified as a potential-field extrapolation from $B_z$ in the uppermost cell within the simulation domain. Time-dependent bottom boundary conditions are imposed for both the velocity and magnetic fields, which is extracted from the solar convective dynamo simulation by \citet{Fan2014}, whereas the thermal quantities at the bottom boundary are fully governed by the default MURaM setup. The code uses explicit numerical resistivity and viscosity as described in \citet{Rempel2014}. The dissipations are highly inhomogeneous in time and space, which are strongly suppressed in smooth regions and enhanced in regions with strong gradient, particularly monotonicity changes. During the calculation, the numerical resistive and viscose dissipations are added in the internal energy such that the total energy is conserved. Further details of the simulation setup can be found in \citet{Chen2017,Chen2022}.

\begin{figure*}[ht!]
\centering
\includegraphics[width=\textwidth]{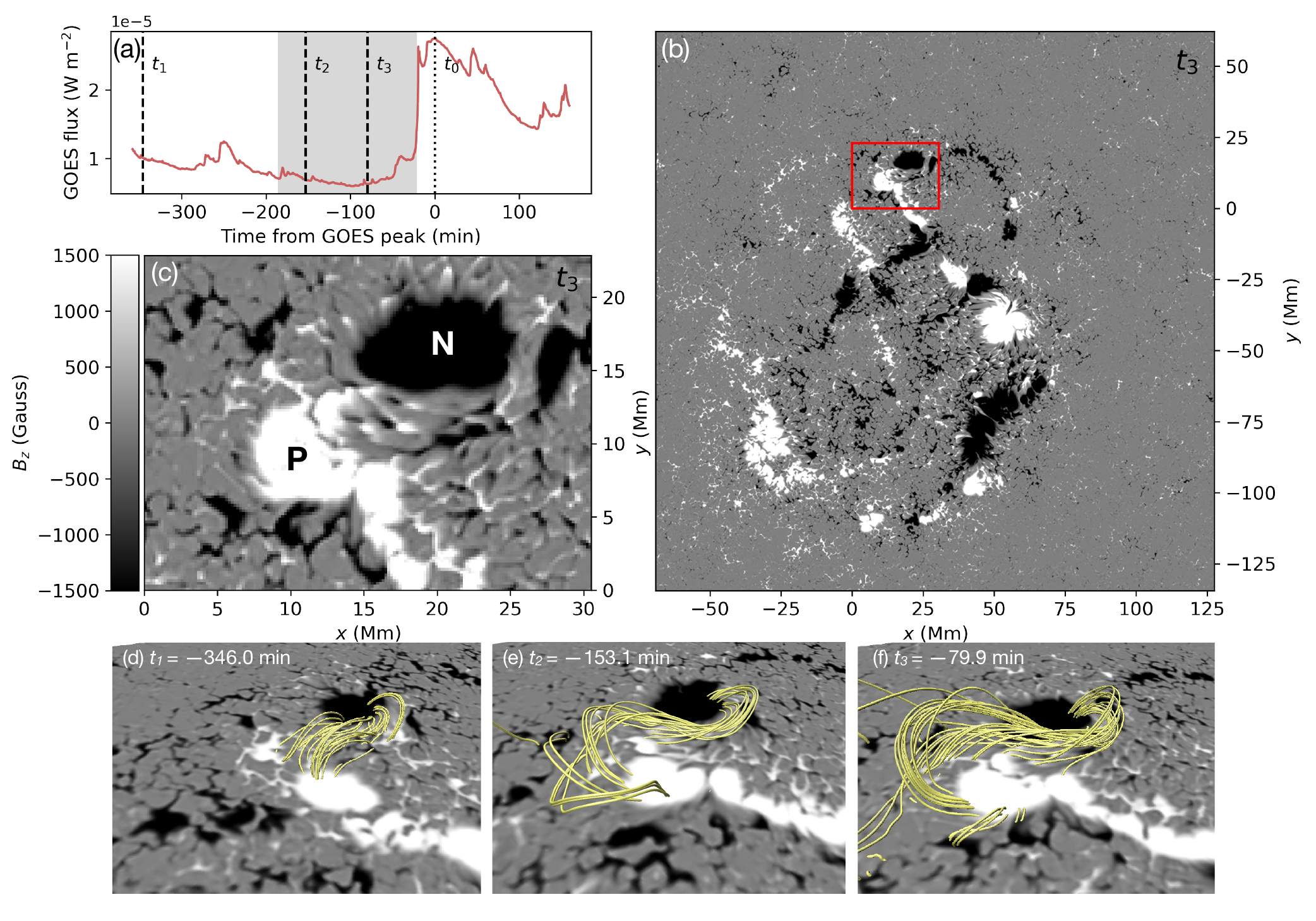} 
\caption{Overview of the magnetic configuration. (a) Synthetic GOES light curve of the simulation. The eruption associated with the flux rope analyzed in this study is characterized by an impulsive enhancement of synthetic GOES flux. We define the time when the GOES flux reaches its maximum during the eruption as $t_0$ = 0 s and mark it with vertical dotted line. The gray-shaded region denotes the time interval considered in the subsequent analysis, covering the period from the initial stage of flux rope formation to the onset of its eruption. Other representative snapshots of the evolution ($t_1$, $t_2$, and $t_3$) are indicated by vertical dashed lines. (b) Photospheric magnetogram of the full simulation domain, with the core region of the eruption highlighted by a red box. (c) Zoom-in view of the core region, where P and N denote the main positive and negative sunspots, respectively. (d)-(f) Temporal evolution of the pre-eruptive magnetic structure. The yellow solid tubes illustrate selected magnetic field lines. The 3D visualization is produced using VAPOR \citep{VAPOR}.}\label{fig1}
\end{figure*}

The evolution of the synthetic Geostationary Operational Environmental Satellite (GOES) 1-8 $\text{\AA}$ flux before and during the eruption is shown in Figure \ref{fig1} (a), which releases magnetic energy comparable to that of an M-class flare. The period of interest in the following analysis is indicated by the gray-shaded region, which spans the entire formation process of the flux rope prior to its eruption. We define the peak time of the GOES flux as $t_0 = 0$ s, and all subsequent times are given relative to $t_0$. Note that there is an obvious delay between this peak - corresponding to the time when evaporation flows fill the post flare loops - and the onset of the eruption, which appears as a slightly lower but much sharper spike at $t \sim - 20$ min. In Figure \ref{fig1}(b) we present a snapshot of the magnetogram over the entire simulation domain, where the original location of the flux rope is marked by a red box; a zoom-in view of this selected region is shown in Figure \ref{fig1}(c). Note that all quantitative analyses hereafter are conducted within the selected region.

\section{Results} \label{sec:result}
\subsection{Identifying the Magnetic Flux Rope in Simulation}\label{subsec:identify}
The dominant magnetic feature in the region of interest (Figure \ref{fig1}(c)) is a sunspot pair consisting of a compact, circular negative polarity (marked as N) and an extended positive polarity (marked as P). Several small-scale magnetic patches are also present between and around the main sunspot pair P-N. Figure \ref{fig1}(d)-(f) shows the temporal evolution of the main magnetic structure in the atmosphere. More than five hours before the flare peak, only magnetic arcades are present along the sunspot pair (Figure \ref{fig1}(d)); later, a seed of flux rope appears and gradually develops into a well-structured pre-eruptive flux rope, as shown in Figure \ref{fig1}(f).

Identifying the volume occupied by the flux rope is crucial for a quantitative investigation of its formation. Based on its qualitative identification, we adopt the twist number $\mathcal{T}_w$ and the squashing factor $Q$ to identify the flux rope, both of which have been widely used in previous studies \citep{Liu2016,Masson2017,Duan2019}. $\mathcal{T}_w$ (see Equations (14) and (16) of \citet{Berger2006}) quantifies the number of turns that neighboring magnetic field lines wind around each other and exhibits a coherent distribution with larger values within a flux rope. Meanwhile, $Q$ (defined as Equations
(11), (12), and (14) of \citet{Titov2007}), which characterizes changes in magnetic connectivity between magnetic field lines , reaches high values in the quasi-separatrix layer (QSL) surrounding the flux rope. The identification is carried out as follows. 

\begin{figure*}[ht!]
\centering
\begin{interactive}{animation}{figure2animation.mp4}
\includegraphics[width=\textwidth]{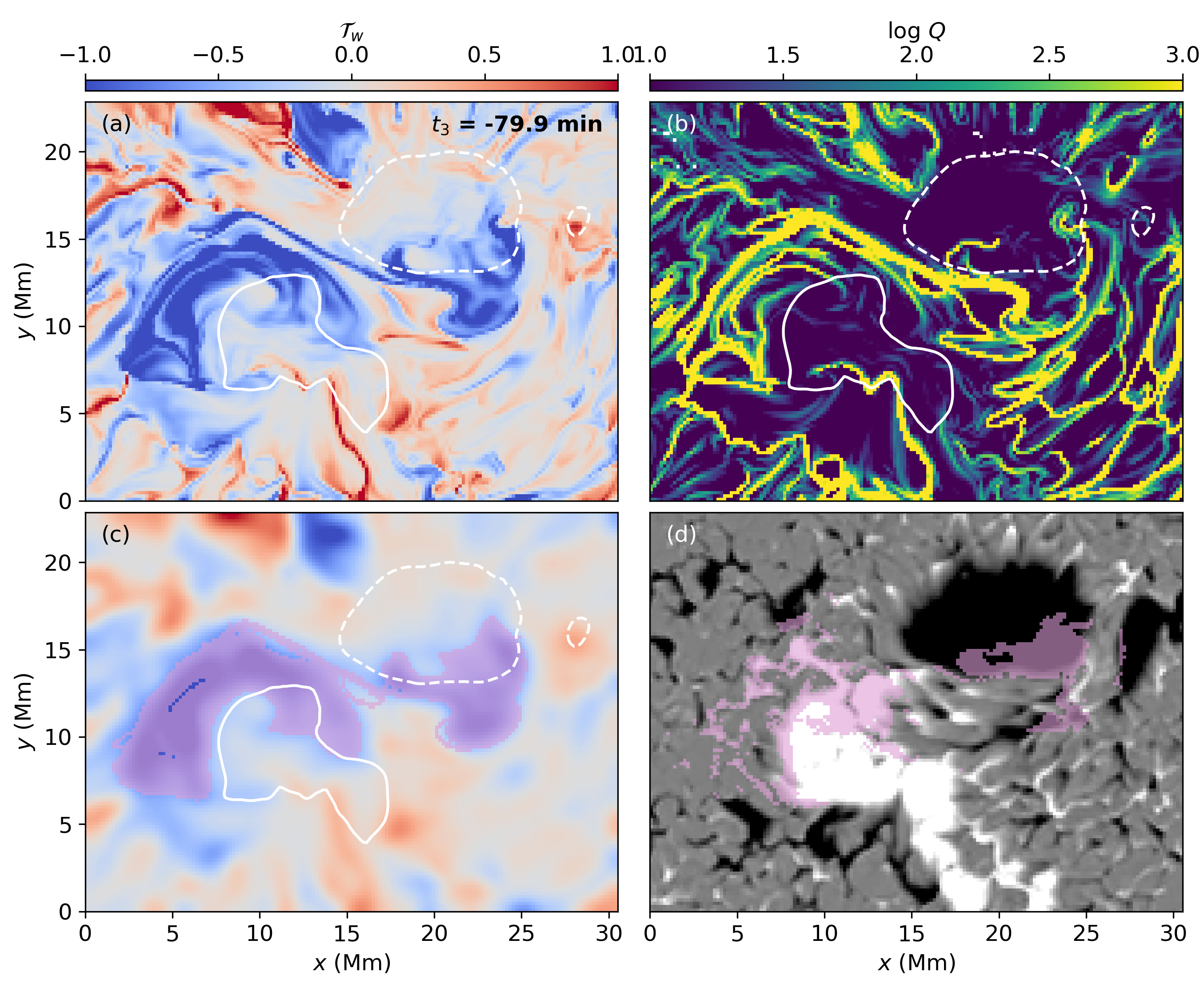}
\end{interactive}
\caption{An example of flux rope identification.
(a)(b) Distributions of $\mathcal{T}_w$ and $\rm{log_{10}}$$\ Q$, respectively, at $t_3 = -$79.9 min on a horizontal plane at $z =$ 0.83 Mm, where the averaged plasma $\beta$ is approximately unity. (c) Smoothed $\mathcal{T}_w$ map overlaid with the cross section of flux rope (pink shading) on this layer identified by the region growing method. The white solid and dashed contours in (a)-(c) outline the regions with $B_z = \pm$1000 Gauss on the $\beta$ unity layer. (d) Photospheric cross section of the flux rope (pink shading) overlaid on the magnetogram. An animation showing the evolution of $\mathcal{T}_w$, $\rm{log_{10}}$$\ Q$, and the cross sections of the flux rope during flux rope formation is available, which spans the simulated time period from $t$ = -186.3 min to $t$ = - 21.0 min, aligning with the interval marked by the gray shaded in Figure \ref{fig1}(a).}\label{fig2}
\end{figure*}

First, we calculate the distribution of $\mathcal{T}_w$ and $Q$ in a horizontal plane at $z_{\beta} = 0.83 ~\mathrm{Mm}$, where the average plasma $\beta$ is approximately unity. These quantities are calculated using the code provided by \citet{Zhang2022}, and an example at $t_3 = -79.9 ~\mathrm{min}$ is shown in Figure \ref{fig2}. The flux rope in this event exhibits a negative twist, and thus its cross section appears as a coherent region of negative $\mathcal{T}_w$ that extends along the edges of the sunspot pair (Figure \ref{fig2}(a)). Simultaneously, it is partly isolated from the background by strong QSL, which corresponds to regions of high $Q$, as indicated by the $Q$ map (Figure \ref{fig2}(b)). 
Then, we delineate the cross section of the flux rope on this layer using the region growing method \citep{Adams1994}, based on the smoothed $\mathcal{T}_w$ map and the contour of strong QSLs where $\log Q$ exceeds a threshold empirically chosen in the range of 2–3.
Prior to $t = -186.3$ min, a continuous negative twist pattern has already developed in the AR, indicating that the magnetic field gradually becomes sheared and twisted. However, no coherent QSL contour is present near the positive sunspot, which implies that a distinct flux rope has not yet formed. At $t = –186.3$ min, a well-defined hook-shaped QSL forms at $x \sim 8$ Mm, $y \sim 11$ Mm (see the accompanying animation of Figure \ref{fig2}). Hereafter, the identified region evolves smoothly and continuously, suggesting the formation of a coherent flux rope. Therefore, we identify $t = –186.3$ min as the onset time for flux rope formation. Finally, by tracing the field lines rooted in the cross section identified on the plane where $z = z_{\beta}$, we reconstruct the 3D structure of the flux rope and its horizontal cross sections at the other altitudes. Given the qualitative features of a flux rope, the reconstructed 3D structure should be independent of the starting point of field line tracing. To validate the consistency of our identification method, we extract the intersection of the identified 3D flux rope with a $yz$-plane to obtain the vertical cross section of flux rope, and find it well consistent with the high $\mathcal{T}_w$ pattern bounded by high $Q$ contours on the same plane, demonstrating the identified flux rope forms a coherent, well-defined structure. We define the horizontal cross section of the flux rope as $S_{\rm MFR}(t,z)$, which varies with time $t$ (in minutes) and altitude $z$ (in Mm), and illustrate $S_{\rm MFR}(t_3,0 \ \mathrm{Mm})$ and $S_{\rm MFR}(t_3,z_{\beta})$ as pink-shaded regions in Figure \ref{fig2} (c) and (d), respectively. We also define the area of the active region $S_{\rm AR}$ as the rectangular region bounded by $0 \le x \le 31~\mathrm{Mm}$ and $0 \le y \le 23~\mathrm{Mm}$.  

The evolution of the cross sections of the flux rope provides insights into the development of its 3D magnetic topology, characterized by increasing spatial extent and structural coherence. On the plane where $z = z_{\beta}$, the cross section of the flux rope $S_{\rm MFR}(t,z_{\beta})$ gradually contracts prior to $t = -140$ min, indicating that the flux rope becomes increasingly compact. Between $t = -160$ min and $t = -140$ min, the QSL near the edge of the positive sunspot undergoes significant changes, indicating a rapid restructuring of the magnetic connectivity. Following the QSL evolution, the cross section near the positive sunspot abruptly extends toward the negative-$x$ and negative-$y$ directions, suggesting that additional field lines may have been incorporated into the flux rope as a result of the topological changes. From approximately $t = -120$ min, the cross sections near the positive and negative sunspots maintain a relatively coherent morphology while extending laterally in opposite directions, indicating a longitudinal elongation of the flux rope. On the photosphere, the flux rope cross section on the negative-polarity side consistently lies along the lower edge of the main negative sunspot N. On the positive-polarity side, the cross section initially appears fragmented, primarily distributed across several dispersed patches in the upper part of the main positive sunspot P. As the flux rope forms and develops, the positive part of $S_{\rm MFR}(t,0 \ \mathrm{Mm})$ gradually becomes more coherent and concentrates along the upper and outer edge of P.

\subsection{Helicity Flux in the Photosphere}
It is believed that the magnetic helicity injected through the photosphere plays a key role in the accumulation of magnetic free energy in the solar atmosphere. Therefore, we employ the relative magnetic helicity flux as a proxy to characterize the evolution of the flux rope and to quantify the contributions of different formation mechanisms. 

\begin{figure*}
\centering
\begin{interactive}{animation}{figure3animation.mp4}
\includegraphics[width=\textwidth]{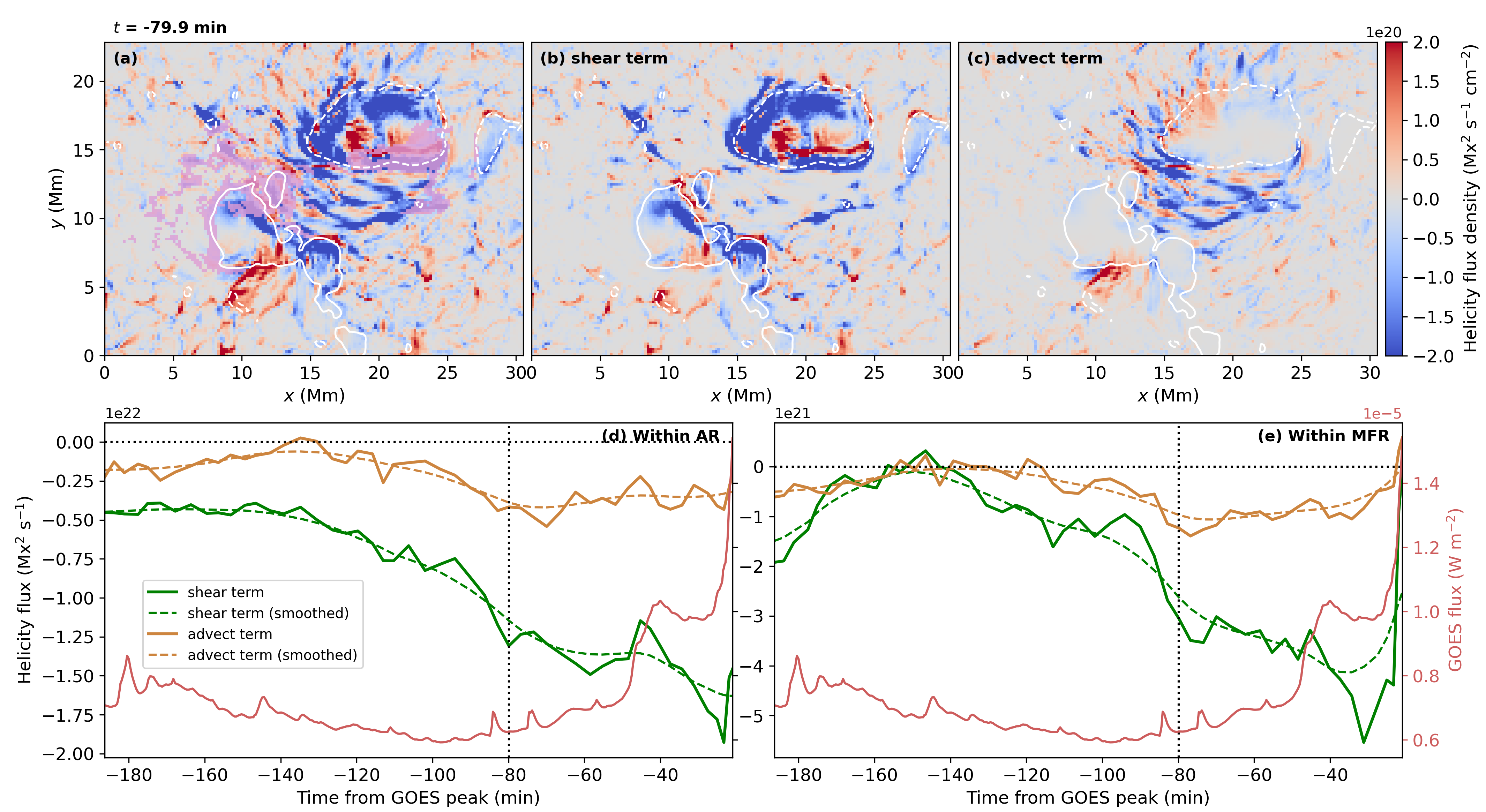} 
\end{interactive}
\caption{Distribution of helicity flux and its components on photosphere where $z = 0$ Mm. (a)-(c) Distributions of helicity flux $\dot{h}$, its shear term $\dot{h}_\mathrm{s}$, and its advection term $\dot{h}_\mathrm{a}$, respectively. White solid and dashed contours indicate regions with photospheric $B_z = \pm$1500 Gauss. The pink-shaded area in (a) marks the cross section of the flux rope on the photosphere. (d) Temporal evolution of the shear term $\dot{H}_\mathrm{s,AR}(t,0 \ \mathrm{Mm})$ and the advection term $\dot{H}_\mathrm{a,AR}(t,0 \ \mathrm{Mm})$ of the helicity flux in the AR, shown as green and orange solid curves, respectively. Dashed curves indicate the corresponding smoothed profiles. Horizontal dotted line marks the zero level, and vertical dotted line denotes the time corresponding to (a)-(c). The red curve represents the synthetic GOES flux. (e) Temporal evolution of the shear term $\dot{H}_\mathrm{s,MFR}(t,0 \ \mathrm{Mm})$ and the advection term $\dot{H}_\mathrm{a,MFR}(t,0 \ \mathrm{Mm})$ of helicity flux in the MFR cross section.Curve styles and symbols are the same as in (d). An animation showing the evolution of helicity flux density on photosphere is available, which spans the simulated time period from $t$ = -186.3 min to $t$ = - 21.0 min.} \label{fig3}
\end{figure*}

The relative magnetic helicity within a volume $\mathcal{V}$ is defined as:
\begin{equation}\label{eq1}
    H = \int_\mathcal{V} (\bm{A} + \bm{A}_{\text{p}}) \cdot (\bm{B} - \bm{B}_{\text{p}}) \, \mathrm{d}^{3}x,
\end{equation}
where $\bm{B}$ and $\bm{A}$ denote the magnetic field and its vector potential, respectively, while the subscript $\mathrm{p}$ indicates quantities associated with the potential field \citep{Finn1985}. For this study, we adopt the Coulomb gauge to calculate the vector potential. Given the low-resistivity nature of solar plasma, the temporal variation of relative magnetic helicity in the volume $\mathcal{V}$, $\text{d}H/\text{d}t$, mainly arises from helicity flux (helicity injection rate) from its boundary, which corresponds to the helicity flux density $\dot{h}$ integrated over the surface $\mathcal{S}$. While the spatial distribution of helicity density is gauge dependent, the integrated helicity flux through the photospheric surface is gauge invariant \citep{Berger1984}, and the quantitative analysis hereafter are only done with surface integrated flux. In most active regions, the magnetic helicity is thought to be injected primarily through the bottom surface, and in observations it is usually evaluated on the photosphere \citep{Pevtsov2014}. The helicity flux (density) can be further decomposed into two distinct physical contributions: the advection term $\dot{H}_\mathrm{a}$ ($\dot{h}_\mathrm{a}$) and the shear term $\dot{H}_\mathrm{s}$ ($\dot{h}_\mathrm{s}$):
\begin{align}
\frac{\text{d}H}{\text{d}t} &= \dot{H}_\mathrm{a} + \dot{H}_\mathrm{s}, \\
\dot{H}_\mathrm{a} &= \int_{\mathcal{S}} \dot{h}_\mathrm{a} \, \text{d}\mathcal{S}, \\
\dot{h}_\mathrm{a} &= 2 (\bm{A}_{\text{p}} \cdot \bm{B}_\mathrm{t}) v_\mathrm{con d,n}, \label{eq:ha} \\
\dot{H}_\mathrm{s} &= \int_{\mathcal{S}} \dot{h}_\mathrm{s} \, \text{d}\mathcal{S}, \\
\dot{h}_\mathrm{s} &= -2 (\bm{A}_{\text{p}} \cdot \bm{v}_\mathrm{cond,t}) B_\mathrm{n} \label{eq:hs}, \\
\dot{h} &= \dot{h}_\mathrm{a}+\dot{h}_\mathrm{s}.
\end{align}
Here, the subscript $\mathrm{t}$ and $\mathrm{n}$ denote the tangential and normal components relative to the bottom surface, respectively, and the velocity is restricted to the conductive component perpendicular to the magnetic field:
\begin{align}
\bm{v}_{\rm cond}   &= \bm{v} - (\bm{v} \cdot \hat{\bm{B}}) \, \hat{\bm{B}}, \\
\hat{\bm{B}} &= \bm{B}/B,
\end{align}
where $\bm{v}$ and $\bm{B}$ are the plasma velocity and magnetic field in Cartesian coordinates, and $\hat{\bm{B}}$ its unit vector of magnetic field.
It is obvious that the advection term of helicity flux density on the photosphere $\dot{h}_\mathrm{a}(t,0 \ \mathrm{Mm})$ arises from the vertical motions of horizontal magnetic field and represents the contribution from flux emergence and submergence; meanwhile, the shear term on the photosphere $\dot{h}_\mathrm{s}(t,0 \ \mathrm{Mm})$ is carried by horizontal motions of vertical magnetic field and reflects the effect of the flux cancellation mechanism \citep{Pariat2005}. Figure \ref{fig3}(a)-(c) display the distributions of the photospheric helicity flux density $\dot{h}(t,0 \ \mathrm{Mm})$, its shear term $\dot{h}_\mathrm{s}(t,0 \ \mathrm{Mm})$, and advection term $\dot{h}_\mathrm{a}(t,0 \ \mathrm{Mm})$, respectively. The cross section of the flux rope $S_{\rm MFR}(t_3,0 \ \mathrm{Mm})$ is indicated by the pink shadow in Figure \ref{fig3}(a). 

To quantify the relative roles of flux emergence and photospheric motions, we integrate the advection and shear terms of the helicity flux density over the entire active region $S_{\rm AR}$ and restricted to the flux rope cross section $S_{\rm MFR}$, respectively. The shear and advection terms of helicity flux in the active region, $\dot{H}_\mathrm{s,AR}(t,0 \ \mathrm{Mm}) = \int_{S_\mathrm{AR}} \dot{h}_\mathrm{s}(t,0 \ \mathrm{Mm}) \, \text{d}\mathcal{S}$ and  $\dot{H}_\mathrm{a,AR}(t,0 \ \mathrm{Mm}) = \int_{S_\mathrm{AR}} \dot{h}_\mathrm{a}(t,0 \ \mathrm{Mm}) \, \text{d}\mathcal{S}$, are shown as green and orange curves in Figure \ref{fig3}(d). Similarly, the corresponding terms in the flux rope cross section, $\dot{H}_\mathrm{s,MFR}(t,0 \ \mathrm{Mm}) = \int_{S_\mathrm{MFR}(t,0 \ \mathrm{Mm})} \dot{h}_\mathrm{s}(t,0 \ \mathrm{Mm}) \, \text{d}\mathcal{S}$ and $\dot{H}_\mathrm{a,MFR}(t,0 \ \mathrm{Mm}) = \int_{S_\mathrm{MFR}(t,0 \ \mathrm{Mm})} \dot{h}_\mathrm{a}(t,0 \ \mathrm{Mm}) \, \text{d}\mathcal{S}$, are shown in Figure \ref{fig3}(e). In both regions, negative helicity accumulates during the build-up of the flux rope, and their temporal evolutions exhibit similar trends, reflecting a coupling between the large-scale AR evolution and the localized MFR formation. Since our primary interest lies in the flux rope formation process, we mainly focus on the results integrated within $S_{\rm MFR}$, as shown in Figure \ref{fig3}(e). Before $t = -140$ min, the two terms show a slight decrease. Afterwards, the efficiency of helicity injection carried by both terms increases steadily until the eruption, with the shear term becoming particularly dominant. These results indicate that both flux emergence and photospheric horizontal motions contribute to helicity injection, yet the latter, which corresponds to the flux cancellation mechanism, plays a more significant role in flux rope formation, especially in the phase just before eruption.

\begin{figure*}
\centering
\includegraphics[width=\textwidth]{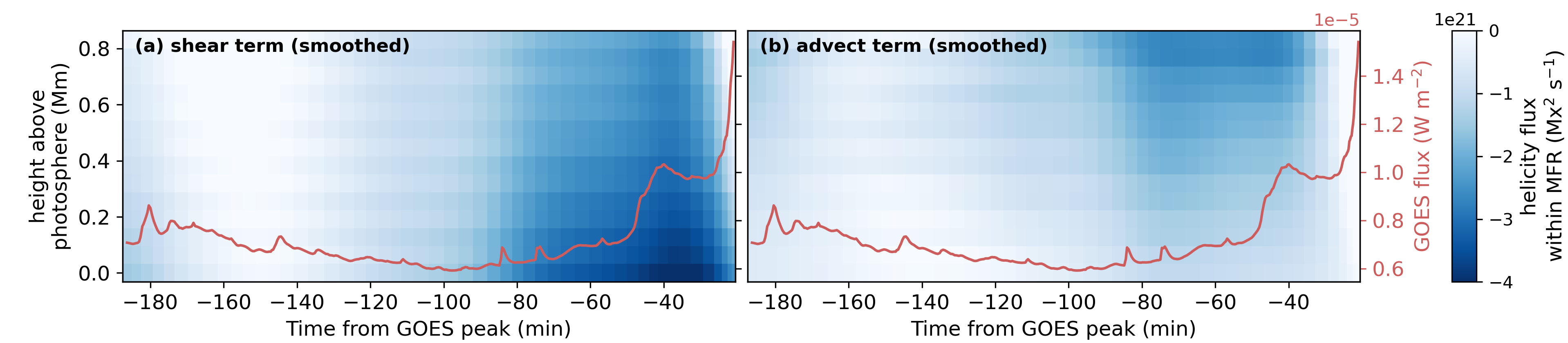} 
\caption{Time–height diagrams of helicity flux injected into the flux rope. Panel (a) shows the evolution of shear term $\dot{H}_\mathrm{s,MFR}(t,z)$, while panel (b) shows the evolution of the advection term $\dot{H}_\mathrm{a,MFR}(t,z)$. The red curves represent the synthetic GOES flux.}\label{fig4}
\end{figure*}

\begin{figure*}[ht!]
\centering
\includegraphics[width=\textwidth]{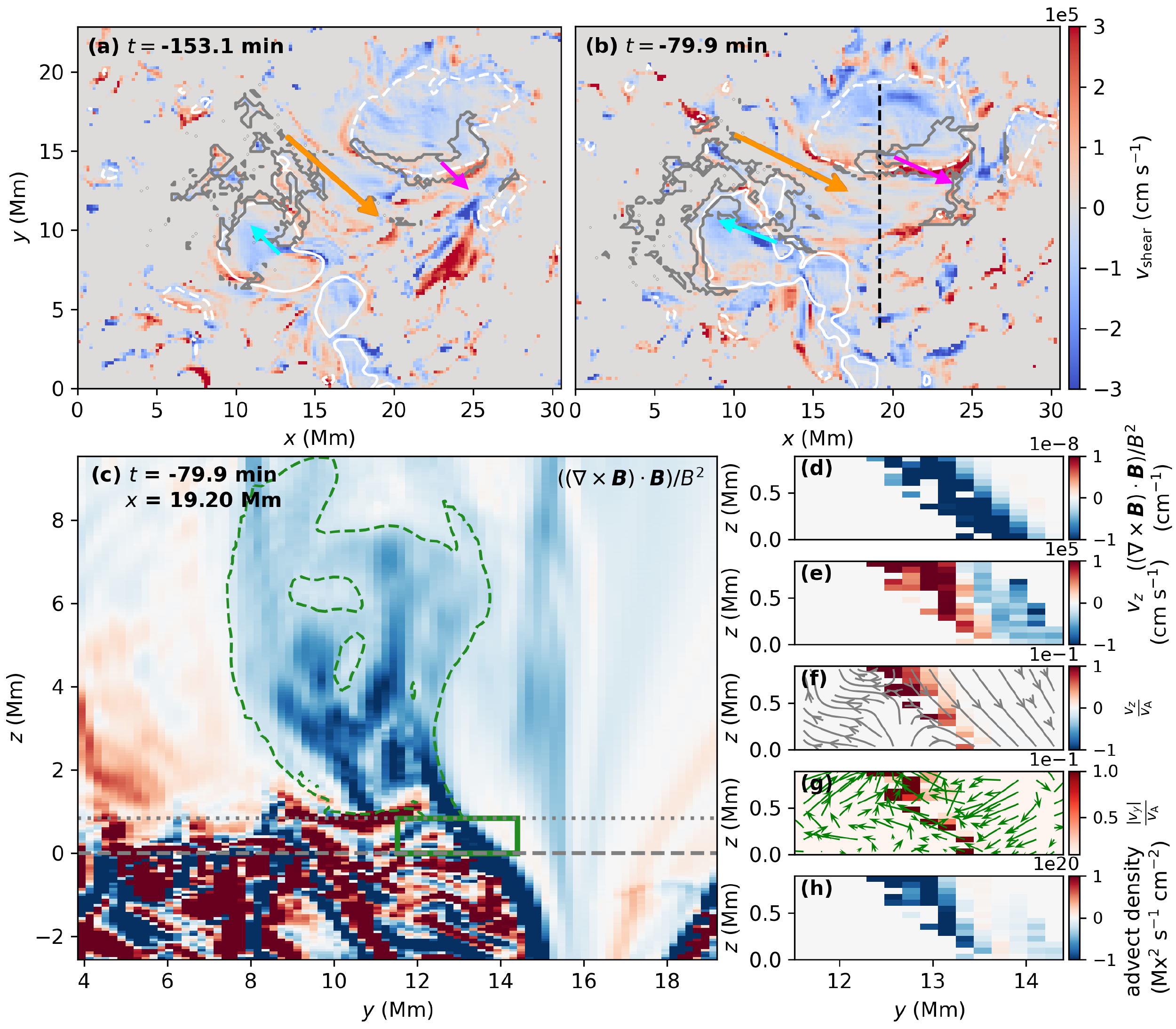} 
\caption{Parameters related to the evolution of helicity flux. (a)(b) Photospheric shearing velocity $v_{\rm sh}$, where a mask based on magnetic field strength is applied to highlight the velocity distribution only within regions of strong magnetic field. The contours of $B_z = \pm$ 1500 Gauss on the photosphere are overplotted in white solid and dashed lines, respectively. The gray contours marks the cross sections of the flux rope on photosphere where $z = 0$ Mm. The orange arrows mark the reference direction for velocity decomposition. The magenta (indigo) arrows denote shearing flows parallel (anti-parallel) to the reference direction. (c) Distribution of $(\nabla \times\bm{B}) \cdot \bm{B}/B^2$ on a $yz$-plane located at $x =$ 19.20 Mm, which is proportional to the field-aligned current density. The dashed and dotted horizontal lines mark the position of $z =$ 0 Mm (photosphere) and $z = 0.83$ Mm, respectively; the height range between them corresponds to that shown in Figure \ref{fig4}. The green dashed contour outlines the approximate location of the flux rope on this plane. The green box encloses the core region of the current sheet, which is selected for a detailed view in (d)-(h). (d)-(h) Zoom-in view of the selected region showing the distribution of $(\nabla \times\bm{B}) \cdot \bm{B}/B^2$, vertical velocity $v_z$, the ratio of $v_z$ to the local Alfven speed $v_{\rm a}$, the ratio of $v_y$ to $v_{\rm a}$, and the density of advection term, respectively. The gray streamlines in (f) depict the orientation of the magnetic field ($B_y$–$B_z$) on this plane, while the green arrows in (g) represent the direction of the in-plane components ($v_y$, $v_z$) of the conductive velocity.}\label{fig5}
\end{figure*}

\subsection{Helicity Flux within the Flux Rope at Different Heights} \label{subsec:height_distribution}
Although photospheric processes, such as flux emergence and horizontal motions, are commonly considered the primary sources of helicity injection into the flux rope, activities associated with flux rope formation, such as magnetic reconnection between sheared magnetic fields, can also occur above the photosphere, potentially redistributing the shear and advection terms at higher altitudes. Therefore, examining the spatial distribution of helicity flux at different heights is crucial for a more complete understanding of the flux rope formation process. 

We integrate the shear and advection terms of the helicity flux density over the flux rope cross sections at each height from the photosphere to the chromosphere. The integration is performed for each time instance, yielding the time-height diagram of the shear term of helicity flux injected into the flux rope, $\dot{H}_\mathrm{s,MFR}(t,z) = \int_{S_\mathrm{MFR}(t,z)} \dot{h_\mathrm{s}}(t,z)  \, \text{d}\mathcal{S}$, and the advection term $\dot{H}_\mathrm{a,MFR}(t,z) = \int_{S_\mathrm{MFR}(t,z)} \dot{h_\mathrm{a}}(t,z)  \, \text{d}\mathcal{S}$. The results are shown in Figure \ref{fig4}. Examining the temporal evolution, we find that both the shear term and the advection term increase after $t = -140$ min within the selected height range. However, the height-dependent behaviors of these two terms differ significantly. The shear term decreases with increasing height, which may naturally result from the reduction of magnetic field strength along height. In contrast, the advection term increases with height, suggesting the presence of additional sources of horizontal magnetic field or upward plasma motions above the photosphere.

To investigate the underlying causes of the temporal and height-dependent variations in helicity flux, we first examine the velocity distribution on the photosphere. Since the shear term exhibits a more pronounced enhancement, our analysis primarily focuses on the horizontal velocity. Typically, the bidirectional shearing flows along the PIL, which drive the magnetic field away from a potential configuration, are considered important for flux rope formation. Here, since the PIL is fragmented, we select a reference direction based on the QSL between the main sunspot pair, which is usually aligned with the flare ribbons related to magnetic structures that store a substantial amount of magnetic free energy prior to the eruption. The shearing velocity $v_{\rm sh}$ is then defined as the component of the horizontal conductive velocity projected onto the reference direction:
\begin{align}
v_{\rm sh}  &= \bm{v}_{\rm cond,h} \cdot \hat{\bm{e}}_{\rm ref}, \\
\bm{v}_{\rm cond,h} &= (v_{{\rm cond},x}, v_{{\rm cond},y}, 0). 
\end{align}
$\hat{\bm{e}}_{\rm ref}$ is the unit vector along the reference direction defined by the QSL between the main sunspot pair, restricted to the $xy$-plane. Figure \ref{fig5}(a) and (b) show the distribution of $v_{\rm sh}$ at two representative snapshots. From $t = -153.1$ min to $t = -79.9$ min, the bidirectional shearing flow at the edges of the main sunspot pair becomes stronger and more extended, as indicated by the magenta and indigo arrows, especially that along the negative sunspot. Moreover, the location of the enhanced shearing flow coincides well with the negative shear term of helicity flux density within the flux rope cross section (Figure \ref{fig3}(b)). Therefore, the enhancement of shearing velocity prior to the eruption can well account for the evolution of the shear term of helicity flux. We also examine the converging velocity perpendicular to the reference direction, but no evident enhancement is detected. This behavior may result from the magnetic configuration of the active region, in which the positive and negative sunspots remained separated during flux rope formation.

The increase of advection term with height contrasts with the typical decay of magnetic field strength at higher altitude. However, this behavior may be attributed to magnetic reconnection occurring above the photosphere, which alters the local magnetic topology, generates horizontal magnetic fields combined with upward outflows, and thus provides an additional source of the advection term. To diagnose the possible occurrence of magnetic reconnection, we analyze the two-dimensional (2D) distribution of current density and velocity on a selected $yz$-plane, the location of which is marked by vertical dashed lines in Figure \ref{fig5}(b). Figure \ref{fig5}(c) presents the distribution of $(\nabla \times\bm{B}) \cdot \bm{B}/B^2$, which represents the normalized current density parallel to the magnetic field and is often enhanced in regions of sharp magnetic field variation. The position of flux rope on this plane is roughly identified by $\mathcal{T}_w$ (the green dashed contour). In particular, a sheet-like structure with enhanced field-aligned current density appears near the lower boundary of the flux rope, indicating the presence of a current sheet. We highlight its core region with a green box and provide a zoom-in view of current density characters, velocity field, and advection term of helicity flux density in the highlighted region in Figure \ref{fig5}(d)-(h).

In Figure \ref{fig5}(e), we identify strong upward velocity at $y \approx 12.4$--$13.2\,\mathrm{Mm}$, coinciding well with the upper part of the current sheet; moreover, as shown in Figure \ref{fig5}(f), its magnitude typically reaches $\sim 0.1$ of the local Alfven speed $v_{A} = {\lvert B_z \rvert}/{\sqrt{4\pi\rho}}$, with peak amplitudes reaching up to $v_A$. Here, only the $B_z$ component is taken into account in the calculation of $v_A$, as it represents the primary magnetic field component involved in magnetic reconnection in the inflow region. The reduction of local outflow speed in some pixels relative to Alfven speed is likely a result of numerical dissipation, which can broden the current sheet and moderate the outflow velocity. The same panel also displays the magnetic field projected onto the $yz$-plane, which resembles the typical topology of magnetic reconnection, with the outflow region well aligned with the rapid upflows. We also examined signatures of reconnection inflows. In Figure \ref{fig5}(g), the horizontal velocity component $v_y$ is compared with $v_A$, with overlaid arrows indicating the in-plane velocity components ($v_y$–$v_z$). Away from the current sheet, large-scale flows converge toward it from both sides, while near and within the current sheet the velocity vectors appear disordered. Nevertheless, the magnitude of $v_y$ reaches $\sim 0.1\,v_A$ within the current sheet, indicating that these flows are likely associated with reconnection inflows. Finally, comparing the locations of the current sheet, the strong upward outflow, and the negative advection term (Figure \ref{fig5}(h)), we find a clear spatial correspondence. Taken together, these results provide compelling evidence for magnetic reconnection above the photosphere, which enhances the advection term with height and contributes significantly to flux rope formation. In combination with the enhanced photospheric shearing motions that dominate the horizontal flows in this event, our results support the classical flux cancellation mechanism, which couples photospheric horizontal motions and magnetic reconnection, as the key process responsible for flux rope formation.

\section{Discussion and Summary} \label{sec:discussion}
\subsection{Identifying Magnetic Flux Ropes in a More Realistic Solar Atmosphere}
Quantifying flux rope formation necessitates a robust approach for flux rope identification, a task that has remained challenging for several years. In observations, the identification of flux ropes usually relies on their indicators. For example, hot channels, which appear as twisted structures in high-temperature passbands, are interpreted as observational indicators of twisted flux rope \citep{Zhang2012}; the dips at the bottom of flux ropes can trap dense plasma, and filaments are therefore sometimes associated with underlying flux ropes \citep{Mackay2010}; in the face-on perspective, twin dimmings at solar surface and the hooks of flare ribbons are used to identify the footpoints of the flux rope, as they are related to the reduction of thermal pressure within the flux ropes during the expansion \citep{Qiu2017} and the QSL enclosing the flux ropes \citep{Cheng2016}, respectively. In theoretical or data-constrained/driven simulations, quantities characterizing the 3D magnetic configuration, such as $\mathcal{T}_w$ and $Q$, are often employed. 

Although all the methods mentioned above help to constrain the location of flux rope, precise determination its boundaries is far from trivial. For instance, the hooks of flare ribbons or QSL are sometimes not fully closed, making it difficult to strictly separate the flux rope from the surrounding magnetic structures; while empirical thresholds are applied for dimmings and twist numbers, due to the varying magnetic and plasma environments, different events may not share a universal threshold. In our work, based on the distribution of $\mathcal{T}_w$ and $Q$, we adopt the region growing algorithm to identify the flux rope. This method, previously successfully applied for MFR footpoint identification \citep{Xing2020}, minimizes the arbitrariness of manual thresholding and provides a reliable delineation of the flux rope region, enabling quantitative investigations of flux rope formation.

\subsection{Relative Importance of the Shear and Emergence}
Quantitative assessment of flux rope formation aims to evaluate the relative contributions of the flux emergence and the surface flow and reconnection processes. Efforts have been made to link measurable signatures with the underlying formation mechanisms. For example, \citet{Zheng2020} analyzed the evolution of magnetic flux and magnetograms during the lifetime of a twisted structure in EUV images, noting that both flux emergence and flux cancellation are crucial for flux rope formation; by carefully examining the magnetogram, photospheric velocity map, small scale brightenings, and further combining observations with magnetic field extrapolations, \citet{Liu2019} ruled out the bodily emergence scenario. Instead, they suggested that flux emergence mainly provides a favorable magnetic environment, whereas shear and reconnection between nonconjugated polarities serve as the direct drivers of flux rope formation. However, previous studies mainly provide qualitative insights, as it is challenging to separately quantify the contributions of the flux emergence and the surface flow and reconnection process, which often occur simultaneously, and to directly compare observational quantities corresponding to each mechanism given their differing physical dimensions. 

Studies of active regions have provided valuable inspiration for our work. Using magnetic helicity flux \citep{Liu2012,Fuentes2024,Sun2024} or winding flux \citep{MacTaggart2021}, prior investigations quantified the relative contributions of normal and tangential flows on photosphere to the evolution and activity of active regions. However, as these studies typically focus on the entire active region, the information specific to the flux rope may be overwhelmed; meanwhile, with only photospheric signatures, the role of magnetic reconnection occurring at higher layers can only be inferred indirectly through complementary evidence. 

Based on aforementioned approaches, we link the advection and shear term of helicity flux directly to flux emergence and horizontal motion. Note that no well-organized twisted flux tubes are observed in the convection zone, which rules out the bodily emergence scenario, and the `emergence' here refers to the upward transport of subsurface magnetic structures that already possess twisted characteristics. Moreover, we minimize contamination from unrelated regions of the AR by restricting the integration area to the flux rope cross sections, thereby achieving a precise quantitative description of the contributions from different mechanisms. Our results demonstrate that the horizontal motions, which are dominated by shearing velocity in this event, plays a more important role in helicity injection and flux rope formation. Furthermore, by investigating the height dependence of helicity flux, we establish a correspondence between magnetic reconnection and helicity variations, providing more complete chain of evidence for the flux cancellation mechanism.

\subsection{Properties of Magnetic Reconnection Involved in Flux Rope Formation}
In Section \ref{subsec:height_distribution}, we present the unambiguous current sheet accompanied by Alfvenic outflows, which occurs as a result of the numerical resistivity. Beyond this macroscopic picture, recent high-resolution 3D simulations by \citet{Wu2025} reveals that such current sheets can host fine structures identified as mini flux ropes, which act as efficient particle accelerators. The fine structures are regarded as products of tearing-mode instability, which develops when the Lundquist number $S \ge 1.0 \times 10^{4}$. The Lundquist number is defined as $S = l_0v_{A}/\eta$, where $l_0$ denotes the characteristic length of current sheet and $\eta$ represents the resistivity. Adopting $l_0 = 1 \ \rm{Mm}$, plasma density $\rho = 1 \times 10^{-10} \ \rm{g \ cm^{-3}}$, magnetic field strength $B_z = 500\  \rm{Gauss}$ near the reconnection region, and an effective numerical resistivity $\eta = 2.9 \times 10^{11} \ \rm{cm^2 \ s^{-1}}$ \citep{Rempel2017}, the corresponding Lundquist number $S$ is estimated to be $4.9 \times 10^{3}$, which lies below the theoretical threshold for tearing-mode instability. Meanwhile, the size of the current sheet here is relatively small and it spans only a limited number of grid points during the period we focus on, which is insufficient to capture any potential fine structures. Therefore, higher resolution is required to resolve the mini flux ropes within the current sheet. For instance, while our simulation employs a uniform horizontal grid spacing of approximately 192 km, the work by \citet{Wu2025} achieves a local resolution of approximately 52 km within the current sheet, which is about four times finer than ours. Other 3D simulations that successfully resolve the fine structures also utilize grid spacings on the order of tens of kilometers \citep{Jiang2021,Wangyl2023}. Implementing such high resolution uniformly across our global-scale computational domain is computationally prohibitive, and the Adaptive Mesh Refinement (AMR) would represent a potentially viable technical solution.

\subsection{The Role of Active Region Dynamics in Flux Rope Formation}
In contrast to idealized simulations, the magnetic field in our study is extremely complex, which brings additional challenges for analysis but also provides opportunities for a more comprehensive investigation of flux rope formation, especially for establishing a coherent physical framework connecting active region evolution to flux rope formation.

The example of magnetic reconnection discussed in Session \ref{subsec:height_distribution} is located near the edge of the main negative sunspot, where strong photospheric shearing motion is present. In our simulation incorporating the flux emergence process, such shearing motion arises self-consistently as a result of active region's evolution. As the twisted flux tube emerges from the convection zone into the atmosphere, it expands naturally due to the decrease in pressure, generating gradients in both magnetic field strength and local twist density along the magnetic field line across the photosphere. The gradient in horizontal magnetic field can produce oppositely directed magnetic tension forces at the two footpoints of individual field lines, driving bidirectional shear flows \citep{Manchester2004,Manchester2007,Fang2010}, while the gradient in local twist density induces a net axis torque, leading to rotational motions of the flux tube \citep{Longcope2000,Fan2009}. Both effects contribute to the shearing velocity at the sunspots edge, facilitating the transport of helicity and free energy from the convection zone to the atmosphere and providing favorable conditions for subsequent magnetic reconnection.

During the formation process, the sunspot pair where the flux rope is primarily anchored remains well separated, accompanied by continuous flux emergence and ongoing development of the active region. Consequently, it is challenging to measure the amount of canceled magnetic flux associated with flux rope formation and to compare with cases that meet the scenarios of classical flux cancellation model, such as idealized simulations that assume adjacent bipolar configurations without flux emergence \citep{Aulanier2010,Xing2024a}, or observational events occurring between nonconjugated polarities that collide with each other \citep{Liu2019,GuoJH2024}. Nevertheless, the underlying physical mechanism remains consistent, involving photospheric horizontal motions followed by magnetic reconnection.

Meanwhile, we observe several nearby small magnetic elements also undergoing flux cancellation, with part of the flux rope footpoints anchored at these locations (e.g., the mixed-polarity magnetic structure at $x \sim 25$ Mm, $y \sim 12$ Mm in Figure \ref{fig2}(d)). Magnetic field visualization (not shown here) reveals that the pairs of tiny elements that cancel with each other are actually connected to the same magnetic field lines in the convection zone,indicating that they represent distinct footpoints of a single flux tube that partly emerges into the atmosphere in a fragmented manner. Such an emergence scenario on solar surface was first proposed by \citet{Pariat2004}, describing the rise of undulatory flux tubes under Parker instability \citep{Parker1966,Horiuchi1988,Matsumoto1992}. At these locations, the observed flux cancellation may result from two possible scenarios: (1) direct emergence of U-loops, or (2) reconnection above the U-loops followed by submergence of the $\Omega$-loops formed after reconnection. The latter process, which can remove dense plasma from the field line, is considered as a more effective pathway for facilitating magnetic flux emergence \citep{Lites2009,Cheung2010}, and can also contribute to the advection term of helicity flux above the photosphere. However, in this event, since the flux rope are primarily anchored at the edges of the main sunspot pair, we argue the reconnection above the U-loops contributes only marginally to the advection term at higher altitudes. 

\begin{figure*}
\centering
\includegraphics[width=\textwidth]{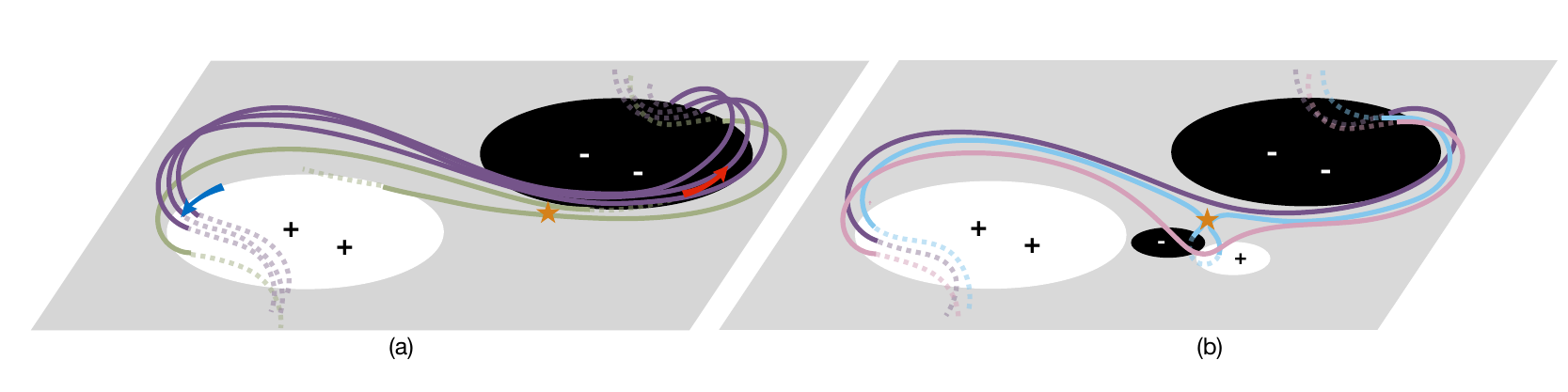} 
\caption{Cartoon illustrating two mechanisms of magnetic flux rope formation. (a) Flux cancellation mechanism that plays a dominant role in flux rope formation, involving horizontal motions and reconnection between distinct field lines. The green lines represent magnetic field lines prior to reconnection, which are of low twist number, while the purple lines depict the flux rope formed through this process. Solid segments denote the parts of field lines above the photosphere, and dotted segments indicate those extending into the convection zone. The orange star marks the reconnection site. The gray plane corresponds to the photosphere. The black and white ellipses denote the main sunspots of negative and positive polarities, respectively, and the red and blue arrows at their edges indicate bidirectional shearing flows at the edges of polarities. (b) Emergence of twisted magnetic field lines. The pink line indicates the smoothly emerged field line, while the light blue line represents the field line emerging in a fragmented manner and undergoing self-reconnection above the photosphere. The smaller polarity pairs illustrate the small-scale magnetic elements associated with flux emergence. Other symbols are the same as in panel (a).}\label{fig6}
\end{figure*}

\vspace{1em}
In summary, we investigate the formation process of the pre-eruptive magnetic flux rope in a RMHD simulation of active region emergence and, for the first time, provide a quantitative assessment of the contribution from different mechanisms. The key idea of this study is illustrated in the cartoon shown in Figure \ref{fig6}. Both (a) the classic flux cancellation mechanism, involving horizontal motions on the photosphere together with magnetic reconnection, and (b) the flux emergence mechanism, contribute to flux rope formation, with the former playing a dominant role. The surface flow and reconnection process mainly represents the interaction between distinct magnetic field lines. The corresponding observational signatures include horizontal motions of polarities, which is dominated by coherent bidirectional shearing flows at the edge of sunspots in our simulation, and localized brightening and plasma acceleration induced by magnetic reconnection, which may be detectable through spectroscopic observations. For polarity pairs that are close to each other, as in the case of nonconjugated polarities, a significant decrease in unsigned magnetic flux may be detected near the PIL. Observational signatures of the emergence mechanism may appear as small-scale flux convergence and cancellation, as well as enhanced activity above the canceled tiny magnetic elements, given that flux emergence is not a smooth pattern but instead involves undulatory field lines emerging segment by segment. However, these signatures are typically more scattered and transient than those in the flux cancellation mechanism.

It is also worth emphasizing that further investigations are still needed for a deeper understanding of flux rope formation. Regarding flux rope identification, our approach relies on the distributions of $\mathcal{T}_w$ and $Q$, with the $\beta$ unity plane serving as a practical reference; while effective, this method may fail to capture portions of flux ropes entirely embedded below this layer. Photospheric diagnostics, such as sunspot scars \citep{Xing2024b}, may provide additional guidance for more complete flux rope identification. For quantifying the formation process, we adopt magnetic helicity flux as a reference measure; likewise, quantities such as winding flux or Poynting flux could also serve as useful diagnostics. Moreover, flux rope formation is a multi-stage process that may continue during the eruption or even initiate after eruption onset \citep{Jiang2021}, representing a regime not addressed in the present study. Recent studies also suggest that, in addition to the role of large-scale magnetic polarities, convective cell dynamics might provide an additional contribution to the evolution of magnetic structure in solar atmosphere \citep{Toriumi2024}, which remains to be quantified. Taken together, these aspects highlight promising directions for our future work.

\begin{acknowledgments}
 We thank the reviewer for his/her valuable comments, which have helped to improve the manuscript. We also thank Hao Wu and Yulei Wang for insightful discussions. This work is supported by National Key R\&D Program of China under grants 2021YFA1600504. C.W. is supported by China Scholarship Concil program No. 202306190187 and Postgraduate Research \& Practice Innovation Program of Jiangsu Province KYCX23\_0118. T.Y. is supported by JSPS KAKENHI grant No. JP21H04492. F.C. is supported by the National Science Foundation of China (NSFC) No. 12422308 and No. 12373054. C. X. acknowledges the support by the NSFC under grant 12403066, the Fundamental Research Funds for the Central Universities under grant 2024300348, and the Jiangsu Funding Program for Excellent Postdoctoral Talent. Z.L. is supported by JSPS Postdoctoral Fellowships for Research in Japan. 
\end{acknowledgments}

\begin{contribution}
C.W. was responsible for analyzing the simulation data and writing the manuscript. T. Y. guided the research methodology. F.C. conducted the simulations. F.C. and M.D.D. initiated the research topic. C.X. and M.D.D. offered important suggestions on the scientific objectives. Z.L. joined the discussions. T.Y., F.C., C.X., M.D.D., and Z.L. edited the manuscript.
\end{contribution}


\bibliography{manuscript}{}
\bibliographystyle{aasjournalv7}

\end{document}